# Big Data: Challenges, Opportunities and Realities

# Big Data: Challenges, Opportunities, and Realities

Abhay Kumar Bhadani
*Indian Institute of Technology Delhi, India*

Dhanya Jothimani
*Indian Institute of Technology Delhi, India*

**ABSTRACT**

*With the advent of Internet of Things (IoT) and Web 2.0 technologies, there has been a tremendous growth in the amount of data generated. This chapter emphasizes on the need for big data, technological advancements, tools and techniques being used to process big data are discussed. Technological improvements and limitations of existing storage techniques are also presented. Since, the traditional technologies like Relational Database Management System (RDBMS) have their own limitations to handle big data, new technologies have been developed to handle them and to derive useful insights. This chapter presents an overview of big data analytics, its application, advantages, and limitations. Few research issues and future directions are presented in this chapter.*

Keywords: Big Data, Big Data Analytics, Cloud Computing, Data Value Chain, Grid Computing, Hadoop, High Dimensional Data, MapReduce

**INTRODUCTION**

With the digitization of most of the processes, emergence of different social network platforms, blogs, deployment of different kind of sensors, adoption of hand-held digital devices, wearable devices and explosion in the usage of Internet, huge amount of data are being generated on continuous basis. No one can deny that Internet has changed the way businesses operate, functioning of the government, education and lifestyle of people around the world.  Today, this trend is in a transformative stage, where the rate of data generation is very high and the type of data being generated surpasses the capability of existing data storage techniques.  It cannot be denied that these data carry a lot more information than ever before due to the emergence and adoption of Internet.

Over the past two decades, there is a tremendous growth in data. This trend can be observed in almost every field. According to a report by International Data Corporation (IDC), a research company claims that between 2012 and 2020, the amount of information in the digital universe will grow by 35 trillion gigabytes (1 gigabyte equivalent to 40 (four-drawer) file cabinets of text, or two music CDs). That's on par with the number of stars in the physical universe! (Forsyth, 2012).

In the mid-2000s, the emergence of social media, cloud computing, and processing power (through multi-core processors and GPUs) contributed to the rise of big data (Manovich, 2011; Agneeswaran, 2012). As of December 2015, Facebook has an average of 1.04 billion daily active users, 934 million mobile daily active users, available in 70 languages, 125 billion friend connections, 205 billion photos uploaded every day 30 billion pieces of content, 2.7 billion likes, and comments are being posted and 130 average number of friends per Facebook user (Facebook, 2015). This has created new pathways to study social and cultural dynamics.

Making sense out of the vast data can help the organization in informed decision-making and provide competitive advantage. Earlier, organizations used transaction-processing systems that inherently used Relational Data Base Management Systems (RDBMS) and simple data analysis techniques like Structured Query Language (SQL) for their day-to-day operations that helped them in their decision making and planning. However, due to the increase in the size of data especially the unstructured form of data (For example, customer reviews of their Facebook pages or tweets), it has become almost impossible to process these data with the existing storage techniques and plain queries.

In this chapter, an overview of big data ranging from its sources to dimensions is given. The limitations of existing data processing approaches; need for big data analytics and development of new approaches for storing and processing big data are briefed. The set of activities ranging from data generation to data analysis, generally termed as Big Data Value Chain, is discussed followed by various applications of big data analytics. The chapter concludes by discussing the limitations of big data analytics and provides direction to open issues for further research.

## BACKGROUND AND NEED FOR BIG DATA ANALYTICS

Storage and retrieval of vast amount of structured as well as unstructured data at a desirable time lag is a challenge. Some of these limitations to handle and process vast amount of data with the traditional storage techniques led to the emergence of the term Big Data. Though big data has gained attention due to the emergence of the Internet, but it cannot be compared with it. It is beyond the Internet, though, Web makes it easier to collect and share knowledge as well data in raw form. Big Data is about how these data can be stored, processed, and comprehended such that it can be used for predicting the future course of action with a great precision and acceptable time delay.

Marketers focus on target marketing, insurance providers focus on providing personalized insurances to their customers, and healthcare providers focus on providing quality and low-cost treatment to patients. Despite the advancements in data storage, collection, analysis and algorithms related to predicting human behavior; it is important to understand the underlying driving as well as the regulating factors (market, law, social norms and architecture), which can help in developing robust models that can handle big data and yet yield high prediction accuracy (Boyd and Crawford, 2011).

The current and emerging focus of big data analytics is to explore traditional techniques such as rule-based systems, pattern mining, decision trees and other data mining techniques to develop business rules even on the large data sets efficiently. It can be achieved by either developing algorithms that uses distributed data storage, in-memory computation or by using cluster computing for parallel computation. Earlier these processes were carried out using grid computing, which was overtaken by cloud computing in recent days.

## Grid Computing

Grid computing is a means of allocating the computing power in a distributed manner to solve problems that are typically vast and requires lots of computational time and power. It works on the principle of voluntary basis, where the users share their computing and memory resources to be used by others. In this setting, the goal is to access computers only when needed and to scale the problems in such a manner that even small computers can make a contribution to the grid. Every computer that is connected to the Internet and wants to become a part of the grid is considered to be a node in an extremely large computing machine. The main advantage of this computing technique is that it offers an opportunity to harness unused computing power. In the grid environment, the problem is divided and distributed to thousands or even millions of computers to obtain a solution in a cost effective manner. There are a number of applications which are using this technology. For example, weather, astronomy, medicine, multi-player gaming, etc. Typically grid computing works on two dominant models: Commercial Model and Social Model (Smith, 2005).

- **Commercial Model:** It works on the principle that this technology can be used for the commercial purpose by creating large processing centers and sell its capabilities to the users on hourly basis and charge money. The advantage of this model is that Quality of Service (QoS) is maintained and is a trusted way of computation.
- **Social Model:** It works with a concept that these resources should be harnessed for the good of society. Grid computing concept is implemented through software that follows Open Grid Service Architecture (OGSA). Globus toolkit is popular software that implements OGSA and is used in grid computing environment.

The important components of a grid computing system are a Computing Element (CE), a number of Storage Elements (SE) and Worker Nodes (WN). The CE helps to establish a connection with other GRID networks and dispatches jobs on the Worker Nodes using a Workload Management System. The storage of the input and the output of data required for the execution of the job are taken care by the storage element. The Storage Element is in charge with the storage of the input and the output of the data needed for the job execution. Garlasu et al. (2013) proposed a framework for processing big data using grid technologies. By introducing a framework for managing big data, it paves a way to implement it around grid architecture.

Grid computing suffers from several drawbacks, which range from financial, social, legal and regulatory issues. Distributing a commercial project across these volunteered machines also raises the issues related to ownership of the results. Given the aggressiveness with which hackers have created e-mail viruses and cheats for computer games create uninvited problems associated with the benefits associated with the grid computing.

## Cloud Computing

Cloud computing has become a recent buzzword, however, it is not a completely new concept. It has its roots from grid computing and other related areas like utility computing, cluster computing and distributing systems, in general. Cloud computing refers to the concept of computing at a remote location with control at the users' end through a thin client system, computer or even mobile phones. Processing, memory, and storage will be done at the service providers' infrastructure (Figure 1). Users need to connect to the virtual system residing at some remote location, which might run several virtual operating systems on physical servers with the help of virtualization. It supports all sorts of fault tolerant features like live migration, scalable storage, and load balancing (Bhadani & Chaudhary, 2010). It provides the flexibility of scaling the computational power, memory as well as storage dynamically on the fly. It works on "pay as you go" concept and is driven by economies of scale. To some extent, cloud computing is similar to the commercial model of grid computing. Primarily, Cloud computing works on three different levels, namely, *Infrastructure as a Service (IaaS), Platform as a Service (PaaS)* and *Software as a Service (SaaS) (*Bhadani, 2011; Assuncao et al., 2015*)*.

Cloud computing also suffers from similar drawbacks of grid computing like data location, data replication, data segregation, security threats, regulatory compliances, recovery issues, long-term viability, high dependency on the Internet for accessing the remote virtual machine, different laws of different countries, investigative support, etc.

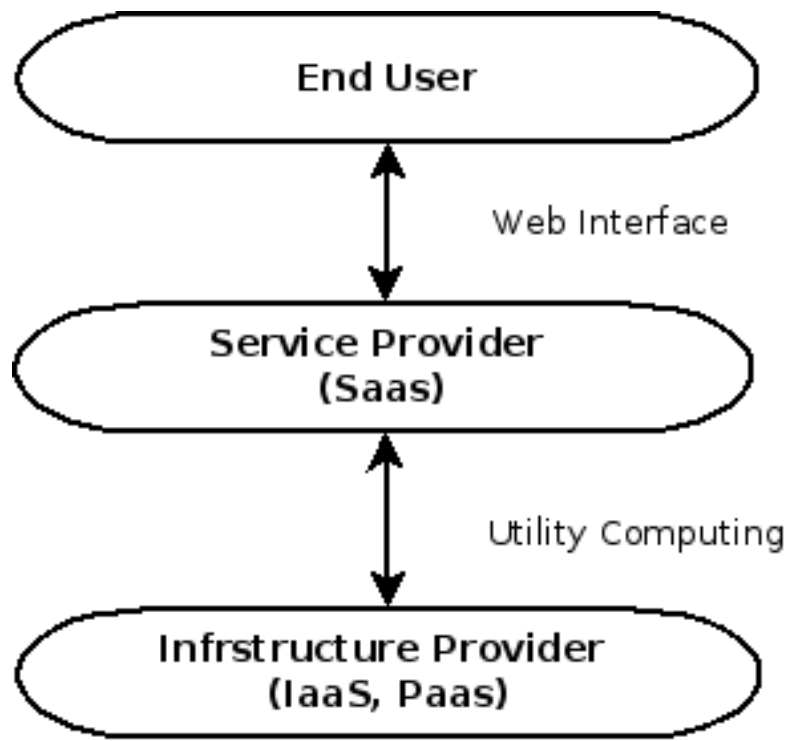

*Figure 1: Basic Model of Cloud Computing*

No technology is full proof, however, given the benefits and drawbacks of grid computing or cloud computing, it might prove useful to process a huge amount of data that need to be processed for big data analysis or live stream data analysis (Foster et al., 2008).

**BIG DATA: DEFINITION, DIMENSIONS, AND SOURCES**

**Definition**

Recently, the word "Big Data" has become a buzzword. It is being used by almost everyone including academicians and industry experts. There are various definitions available in the literature. But the concept of big data dates back to the year 2001, where the challenges of increasing data were addressed with a 3Vs model by Laney (2001). 3Vs, also known as the dimensions of big data, represent the increasing Volume, Variety, and Velocity of data (Assunção et al., 2015). The model was not originally used to define big data but later has been used eventually by various enterprises including Microsoft and IBM to define the same (Meijer, 2011).

In 2010, Apache Hadoop defined big data as "datasets, which could not be captured, managed, and processed by general computers within an acceptable scope" (p.173, Chen et al., 2014). Following this, in 2011, McKinsey Global Institute defined big data as "datasets whose size is beyond the ability of typical database software tools to capture, store, manage, and analyze" p.1 (Manyika et al., 2011). International Data Corporation (IDC) defines "big data technologies as a new generation of technologies and architectures, designed to economically extract value from very large volumes of a wide variety of data, by enabling high-velocity capture, discovery, and/or analysis" (p. 6, Gantz and Reinsel, 2011).

Academicians define big data as huge size of unstructured data produced by high-performance heterogeneous group of applications that spans from social network to scientific computing applications. The datasets range from a few hundred gigabytes to zetabytes that it is beyond the capacity of existing data management tools to capture, store, manage and analyze (Cuzzocrea et al., 2011; Qin et al., 2012; Agneeswaran, 2012).

Though big data has been defined in various forms but there is no specific definition. Few have defined what it does while very few have focused on what it is. The definition of the big data on the basis of 3Vs is relative. What is defined, as big data may not be the same tomorrow? For instance, in future, with the advancements in the storage technologies, the data that is deemed as big data today might be captured. In addition to defining big data, there is a need to understand how to make the best use of this data to obtain valuable information for decision making.

## Dimensions of Big Data

Initially, big data was characterized by the following dimensions, which were, often, referred as 3V model:

a) **Volume:** Volume refers to the magnitude of the data that is being generated and collected. It is increasing at a faster rate from terabytes to petabytes (1024 terabytes) (Zikopoulos et al., 2012; Singh and Singh, 2012). With increase in storage capacities, what cannot be captured and stored now will be possible in future. The classification of big data on the basis of volume is relative with respect to the type of data generated and time. In addition, the type of data, which is often referred as Variety, defines "big" data. Two types of data, for instance, text and video of same volume may require different data management technologies (Gandomi and Haider, 2015).

b) **Velocity:** Velocity refers to the rate of generation of data. Traditional data analytics is based on periodic updates- daily, weekly or monthly. With the increasing rate of data generation, big data should be processed and analyzed in real- or near real-time to make informed decisions. The role of time is very critical here (Singh and Singh, 2012; Gandomi and Haider, 2015). Few domains including Retail, Telecommunications and Finance generate high-frequency data. The data generated through Mobile apps, for instance, demographics, geographical location, and transaction history, can be used in real-time to offer personalized services to the customers. This would help to retain the customers as well as increase the service level.

c) **Variety:** Variety refers to different types of data that are being generated and captured. They extend beyond structured data and fall under the categories of semi-structured and unstructured data (Zikopoulos et al., 2012; Singh and Singh, 2012; Gandomi and Haider, 2015). The data that can be organized using a pre-defined data model are known as structured data. The tabular data in relational databases and Excel are examples of structured data and they constitute only 5% of all existing data (Cukier, 2010). Unstructured data cannot be organized using these pre-defined model and examples include video, text, and audio. Semi-structured data that fall between the categories of structured and unstructured data. Extensible Markup Language (XML) falls under this category.

Later, few more dimensions have been added, which are enumerated below:

d) **Veracity:** Coined by IBM, veracity refers to the unreliability associated with the data sources (Gandomi and Haider, 2015). For instance, sentiment analysis using social media data (Twitter, Facebook, etc.) is subject to uncertainty. There is a need to differentiate the reliable data from uncertain and imprecise data and manage the uncertainty associated with the data.

e) **Variability:** Variability and Complexity were added as additional dimensions by SAS. Often, inconsistency in the big data velocity leads to variation in flow rate of data, which is referred to as variability (Gandomi and Haider, 2015). Data are generated from various sources and there is an increasing complexity in managing data ranging from transactional data to big data. Data generated from different geographical locations have different semantics (Zikopoulos et al., 2012; Forsyth, 2012).

f) **Low-Value density:** Data in its original form is unusable. Data is analyzed to discover very high value (Sun and Heller, 2012). For example, logs from the website cannot be used in its initial form to obtain business value. It must be analyzed to predict the customer behavior.

## Sources of Big Data

Having understood what big data are and their dimensions, here various sources of big data are briefed. Digitization of content by industries is the new source of data (Villars et al., 2011). Advancements in technology also lead to high rate of data generation. For example, one of the biggest surveys in Astronomy, Sloan Digital Sky Survey (SDSS) has recorded a total of 25TB data during their first (2000-2005) and second surveys (2005-2008) combined. With the advancements in the resolution of the telescope, the amount of data collected at the end of their third survey (2008-14) is 100 TB. Use of "smart" instrumentation is another source of big data. Smart meters in the energy sector record the electricity utilization measurement every 15 minutes as compared to monthly readings before.

In addition to social media, Internet of Things (IoT) has, now, become the new source of data. The data can be captured from agriculture, industry, medical care, etc of the smart cities developed based on IoT.

Table 1 summarizes the various types of data produced in different sectors.

*Table 1. Different Sources of Data*

| Sector | Data Produced | Use |
|---|---|---|
| Astronomy | Movement of stars, satellites, etc. | To monitor the activities of asteroid bodies and satellites |
| Financial | News content via video, audio, twitter and news report | To make trading decisions |
| Healthcare | Electronic medical records and images | To aid in short-term public health monitoring and long-term epidemiological research programs |
| Internet of Things (IoT) | Sensor data | To monitor various activities in smart cities |
| Life Sciences | Gene sequences | To analyze genetic variations and potential treatment effectiveness |
| Media/Entertainment | Content and user viewing behavior | To capture more viewers |
| Social Media | Blog posts, tweets, social networking sites, log details | To analyze the customer behavior pattern |
| Telecommunications | Call Detail Records (CDR) | Customer churn management |
| Transportation, Logistics, Retail, Utilities | Sensor data generated from fleet transceivers, RFID tag readers and smart meters | To optimize operations |
| Video Surveillance | Recordings from CCTV to IPTV cameras and recording system | To analyze behavioral patterns for service enhancement and security |

## BIG DATA VALUE CHAIN

Value Chain, the concept introduced by Porter (1980), refers to a set of activities performed by a firm to add value at each step of delivering a product/service to its customers. In a similar way, data value chain refers to the framework that deals with a set of activities to create value from available data. It can be divided into seven phases: data generation, data collection, data transmission, data pre-processing, data storage, data analysis and decision making (Figure 2).

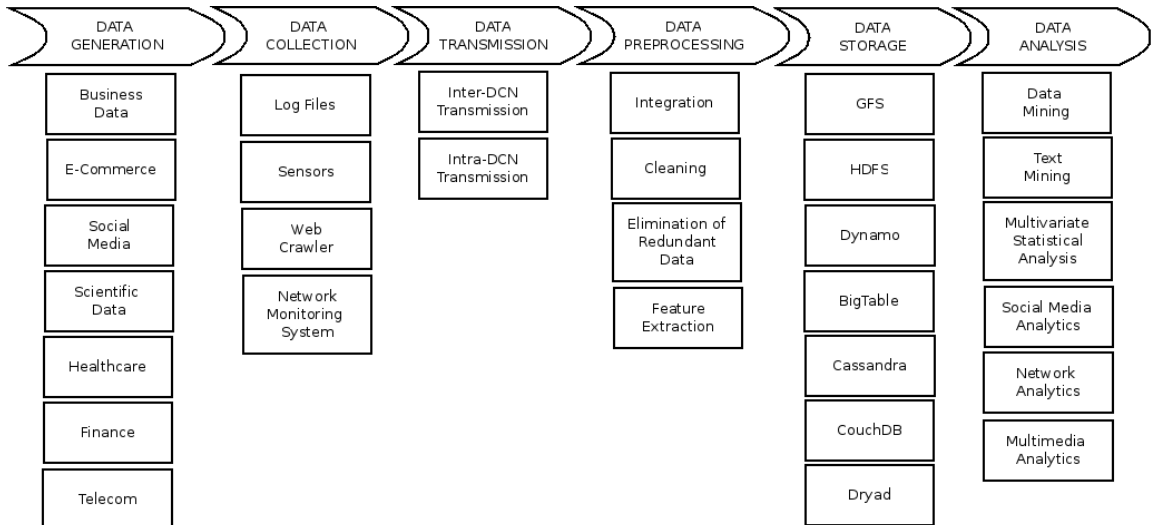

*Figure 2: Big Data Value Chain*

1) **Data Generation:** The first and foremost step the big data value chain is the generation of data. As discussed in the previous section, data is generated from various sources that include data from Call Detail Records (CDR), blogs, Tweets and Facebook Page.

2) **Data Collection:** In this phase, the data is obtained from all possible data sources (Miller and Mork, 2013; Chen et al., 2014). For instance, in order to predict the customer churn in Telecom, data can be obtained from CDRs and opinions/complaints of the customers on Social Networking Sites such as Twitter (in the form of tweets) and Facebook (opinions shared on the company's Facebook page). The most commonly used methods are log files, sensors, web crawlers and network monitoring software (Chen et al., 2014).

3) **Data Transmission:** Once the data is collected, it is transferred to a data storage and processing infrastructure for further processing and analysis. It can be carried out in two phases: Inter-Dynamic Circuit Network (DCN) transmission and Intra-DCN transmissions. Inter-DCN transmission deals with the transfer of data from the data source to the data center while the latter helps in the transfer within the data center. Apart from storage of data, data center helps in collecting, organizing and managing data.

4) **Data Pre-processing:** The data collected from various data sources may be redundant, noisy and inconsistent, hence, in this phase; the data is pre-processed to improve the data quality required for analysis. This also helps to improve the accuracy of the analysis and reduce the storage expenses. The data can be pre-processed with the help of following steps:
    a) Integration: The data from various sources are combined to provide a unified and uniform view of the available data. Data federation and data warehousing are the two commonly used traditional methods. Data warehousing executes the Extract, Transform, and Load (ETL) process. During extract process, the data is selected, collected, processed and analyzed. The process of converting the extracted data to a

standard format is called Transformation process. In Loading, the extracted and transformed data is imported into a storage infrastructure. In order to make data integration dynamic, data can be aggregated from various data sources using a virtual database. It does not contain any data but the details regarding the information related to original data or metadata can be obtained (Miller and Mork, 2013; Chen et al., 2014).

b) Cleaning: The data is checked for accuracy, completeness and consistency. During this process, the data may be deleted and modified to improve the data quality. The general process followed includes following five processes: error types are defined and determined, errors are identified from the data, errors are corrected, error types and corresponding examples are documented, and data entry procedure may be modified to avoid future errors (Maletic and Marcus, 2000).

c) Elimination of Redundant Data: Many datasets have surplus data or data repetitions and are known as data redundancy. This increases the storage cost, leads to data inconsistency and affects the quality of data. In order to overcome this, various data reduction methods such as data filtering and compression, are used. The limitation of these data reduction techniques is that they increase the computational cost. Hence, a cost-benefit analysis should be carried before using data reduction techniques.

5) **Data Storage:** The big data storage systems should provide reliable storage space and powerful access to the data. The distributed storage systems for big data should consider factors like consistency (C), availability (A) and partition tolerance (P). According to the CAP theory proposed by Brewer (2000), the distributed storage systems could meet two requirements simultaneously, that is, either consistency and availability or availability and partition tolerance or consistency and partition tolerance but not all requirements simultaneously (Gilbert and Lynch, 2002). Considerable research is still going on in the area of big data storage mechanism. Little advancement in this respect is Google File System (GFS), Dynamo, BigTable, Cassandra, CouchDB, and Dryad.

6) **Data Analysis:** Once the data is collected, transformed and stored, the next process is data exploitation or data analysis, which is enumerated using the following steps:

   a) Define Metrics: Based on the collected and transformed data, a set of metrics is defined for a particular problem. For instance, to identify a potential customer who is going to churn out, a number of times he/she contacted (be it through a voice call, tweets or complaints on Facebook page) can be considered. (Miller and Mork, 2013).

   b) Select architecture based on analysis type: Based on the timeliness of analysis to be carried out, suitable architecture is selected. Real-time analysis is used in the domain where the data keeps on changing constantly and there is a need for rapid analysis to take actions. Memory-based computations and parallel processing systems are the existing architectures. Fraud detection in retail sectors and telecom fraud are the examples of real-time analysis. The applications that do not require high response time is carried out using offline analysis. The data can be extracted, stored and analyzed relatively later in time. Generally used architecture is Hadoop platform (Chen et al., 2014).

c) Selection of appropriate algorithms and tools: One of the most important steps of data analysis is selection of appropriate techniques for data analysis. Few traditional data analysis techniques like cluster analysis, regression analysis and data mining algorithms, still hold good for big data analytics. Cluster analysis is an unsupervised technique that group's objects based on some features. Data mining techniques help to extract unknown, hidden and useful information from a huge data set. The 10 most powerful data mining algorithms were shortlisted and discussed in Wu et al. (2007). Various tools are available for data analysis including open source softwares and commercial softwares. Few examples of open source softwares are R for data mining and visualization, Weka/Pentaho for machine learning and RapidMiner for machine learning and predictive analysis.

d) Data Visualization: The need for inspecting details at multiple scales and minute details gave rise to data visualization. Visual interfaces along with statistical analyzes and related context help to identify patterns in large data over time (Fisher et al., 2012). Visual Analytics (VA) is defined as "the science of analytical reasoning facilitated by visual interactive interfaces" (Thomas and Cook, 2005). Few visualization tools are Tableau, QlikView, Spotfire, JMP, Jaspersoft, Visual Analytics, Centrifuge, Visual Mining and Board. A comparison of visualization tools based on their data handling functionality, analysis methods and visualization techniques has been discussed in Zhang et al. (2012).

7) **Decision Making:** Based on the analysis and the visualized results, the decision makers can decide whether and how to reward a positive behavior and change a negative one. The details of a particular problem can be analyzed to understand the causes of the problems take informed decisions and plan for necessary actions (Miller and Mork, 2013).

Having discussed about how value can be extracted from big data, an industry regardless of sector should consider three criteria before implementing big data analytics: can useful information be obtained in addition to those obtained from the existing systems, will there be any improvement in the accuracy of information obtained using big data analytics and finally, will implementation of big data analytics help in improving the timeliness of response (Villars et al., 2011).

## TOOLS FOR COLLECTING, PREPROCESSING AND ANALYZING BIG DATA

Advancement in computing architecture is required to handle both the data storage requirements and the heavy server processing required to analyze large volumes and variety of data economically (Villars et al., 2011). This section gives an overview of the technologies adopted to analyze big data.

With the availability of high computational capacity at a relatively inexpensive cost allows researchers to explore the underlying opportunities of big data with the field of data science. Increasingly, data are not solely being captured for record keeping, but to explore them with intelligent systems to derive new insights, which may not have been envisioned at the time of collecting the data. By initiating interesting questions and refining them without experts' intervention, it becomes capable of discovering new information on its own (Dhar, 2013). For instance, if an information can be derived that a specific group of people are prone to cancer and also gives other information such as their diets, daily habits and nature of drugs causing this effect. It would be amazing to develop these kinds of models, which may look like a fiction story at this point of time.

Advancements in these technologies help managers in scenario building analysis. Big data analytics has huge application in various fields including astronomy, healthcare, and telecommunication. Despite advantages, big data analytics has its own limitations and challenges. Security and Privacy issues are main concern for researchers. These advancements as well as limitations resulted in a new interdisciplinary domain called data science, which utilizes the knowledge subjects including psychology, statistics, economics, social science, network science and computer science (Figure 3).

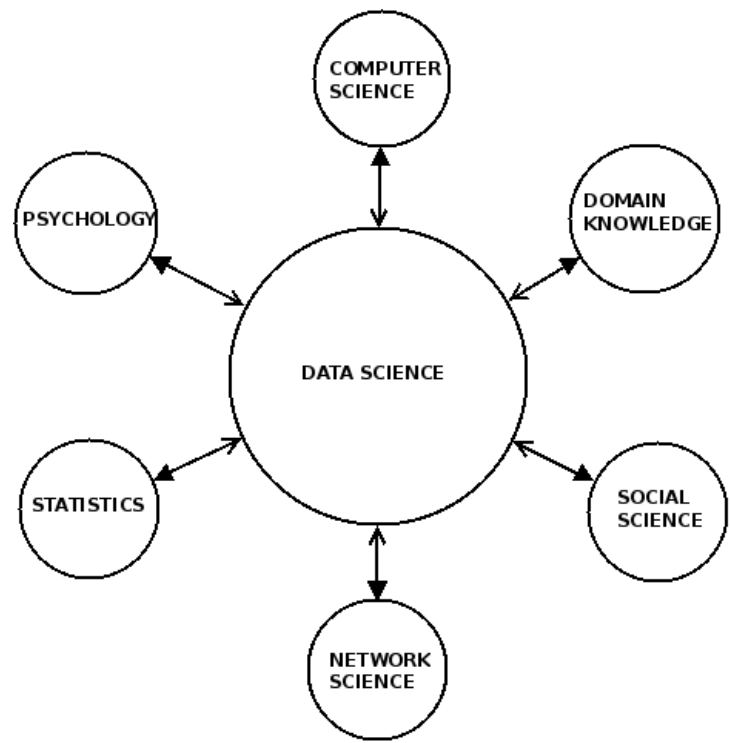

*Figure 3: Evolution of Data Science*

Tools that are being used to collect data encompass various digital devices (for example, mobile devices, camera, wearable devices, and smart watches) and applications that generate enormous data in the form of logs, text, voice, images, and video. In order to process these data, several researchers are coming up with new techniques that help better representation of the unstructured data, which makes sense in big data context to gain useful insights that may not have been envisioned earlier.

## Not only Structured Query Languages (NoSQL)

Relational Database Management System (RDBMS) is the traditional method of managing structured data. RDBMS uses a relational database and schema for storage and retrieval of data. A Data warehouse is used to store and retrieve large datasets. Structured Query Language (SQL) is most commonly used database query language. The data is stored in a data warehouse using dimensional approach and normalized approach (Bakshi, 2012). In dimensional approach, data are divided into fact table and dimension table which supports the fact table. In normalized approach, data is divided into entities creating several tables in a relational database.

Due to Atomicity, Consistency, Isolation and Durability (ACID) constraint, scaling of a large volume of data is not possible. RDBMS is incapable of handling semi-structured and unstructured data (Qin et al., 2012; Mukherjee et al., 2012; Zikopoulos et al., 2012). These limitations of RDBMS led to the concept of NoSQL.

NoSQL stores and manages unstructured data. These databases are also known as "schema-free" databases since they enable quick upgradation of structure of data without table rewrites. NoSQL supports document store, key value stores, BigTable and graph database. It uses looser consistency model than the traditional databases. Data management and data storage functions are separate in NoSQL database (Bakshi, 2012). It allows the scalability of data. Few examples of NoSQL databases are HBase, MangoDB, and Dynamo.

## Hadoop

In 2005, an open source Apache Hadoop project was conceived and implemented on the basis of Google File System and Map Reduce programming paradigm (Prekopcsàk et al., 2011; Bakshi, 2012; Minelli et al., 2013).

### Hadoop Distributed File System (HDFS)

HDFS is the fault-tolerant, scalable, highly configurable distributed storage system for a Hadoop cluster. Data in the Hadoop cluster is broken down into pieces by HDFS and are distributed across different servers in the Hadoop cluster. A small chunk of the whole data set is stored on the server.

### Hadoop MapReduce

MapReduce is a software framework for distributed processing of vast amounts of data in a reliable, fault-tolerant manner. The two distinct phases of MapReduce are:

1) **Map Phase:** In Map phase, the workload is divided into smaller sub-workloads. The tasks are assigned to Mapper, which processes each unit block of data to produce a sorted list of (key, value) pairs. This list, which is the output of mapper, is passed to the next phase. This process is known as shuffling.
2) **Reduce:** In Reduce phase, the input is analyzed and merged to produce the final output which is written to the HDFS in the cluster.

Table 2 summarizes the big data capabilities and the available primary technologies (Sun and Heller, 2012).

*Table 2: Big Data Capabilities and their Primary Technologies*

| Big Data Capability | Primary Technology | Features |
|---|---|---|
| Storage and management capability | Hadoop Distributed File System (HDFS) | Open source distributed file system, Runs on high-performance commodity hardware, Highly scalable storage and automatic data replication |
| Database capability | Oracle NoSQL | Dynamic and flexible schema design, Highly scalable multi-node, multiple data center, fault tolerant, ACID operations, High-performance key-value pair database |
| | Apache HBase | Automatic failover support between Region servers, Automatic and configurable sharding of tables |
| | Apache Cassandra | Fault tolerance capability for every node, Column indexes with the performance of log-structured updates and built-in caching |
| | Apache Hive | Query execution via MapReduce, Uses SQL-like language HiveQL, Easy ETL process either from HDFS or Apache HBase |
| Processing capability | MapReduce | Distribution of data workloads across thousands of nodes, Breaks problem into smaller sub-problems |
| | Apache Hadoop | Highly customizable infrastructure, Highly scalable parallel batch processing, Fault tolerant |
| Data integration capability | Oracle big data connectors, Oracle data integrator | Exports MapReduce results to RDBMS, Hadoop, and other targets, Includes a Graphical User Interface |
| Statistical analysis capability | R and Oracle R Enterprise | Programming language for statistical analysis |

## Limitations of Hadoop

In spite of Hadoop's advantages over RDBMS, it suffers from the following limitations (ParAccel, 2012):
1. **Multiple Copies of Data:** Inefficiency of HDFS leads to creation of multiple copies of the data (minimum 3 copies).
2. **Limited SQL support:** Hadoop offers a limited SQL support and they lack basic functions such as sub-queries, "group by" analytics etc.
3. **Inefficient execution:** Lack of query optimizer leads to inefficient cost-based plan for execution thus resulting in larger cluster compared to similar database.
4. **Challenging Framework:** Complex transformational logic cannot be leveraged using the MapReduce framework.
5. **Lack of Skills:** Knowledge of algorithms and skills for distributed MapReduce development are required for proper implementation.

One of the biggest challenges is to have a computing infrastructure that can analyze high-volume and varied (structured and unstructured) data from multiple sources and to enables real-time analysis of unpredictable content with no apparent schema or structure (Villars et al., 2011).

## SOFTWARE TOOLS FOR HANDLING BIG DATA

There are many tools that help in achieving these goals and help data scientists to process data for analyzing them. Many new languages, frameworks and data storage technologies have emerged that supports handling of big data.

**R:** is an open-source statistical computing language that provides a wide variety of statistical and graphical techniques to derive insights from the data. It has an effective data handling and storage facility and supports vector operations with a suite of operators for faster processing. It has all the features of a standard programming language and supports conditional arguments, loops, and user-defined functions. R is supported by a huge number of packages through Comprehensive R Archive Network (CRAN). It is available on Windows, Linux, and Mac platforms. It has a strong documentation for each package. It has a strong support for data munging, data mining and machine learning algorithms along with a good support for reading and writing in distributed environment, which makes it appropriate for handling big data. However, the memory management, speed, and efficiency are probably the biggest challenge faced by R. R Studio is an Integrated Development Environment that is developed for programming in R language. It is distributed for standalone Desktop machines as well as it supports client-server architecture, which can be accessed from any browser.

**Python:** is yet another popular programming language, which is open source and is supported by Windows, Linux and Mac platforms. It hosts thousands of packages from third-party or community contributed modules. NumPy, Scikit, and Pandas support some of the popular packages for machine learning and data mining for data preprocessing, computing and modeling. NumPy is the base package for scientific computing. It adds support for large, multi-dimensional arrays and matrices with Python. Scikit supports classification, regression, clustering, dimensionality reduction, feature selection, and preprocessing and model selection algorithms.

Pandas help in data munging and preparation for data analysis and modeling. It has strong support for graph analysis with its NetworkX library and nltk for text analytics and Natural language processing. Python is very user-friendly and great for quick and dirty analysis on a problem. It also integrates well with spark through the pyspark library.

**Scala**: is an object-oriented language and has an acronym for "Scalable Language". The object and every operation in Scala is a method-call, just like any object-oriented language. It requires java virtual machine environment. Spark, an in-memory cluster computing framework is written in Scala. Scala is becoming popular programming tool for handling big data problems.

**Apache Spark:** is an in-memory cluster computing technology designed for fast computation, which is implemented in Scala. It uses Hadoop for storage purpose as it has its own cluster management capability. It provides built-in APIs for Java, Scala, and Python. Recently, it has also started supporting R. It comes with 80 high-level operators for interactive querying. The in-memory computation is supported with its Resilient Distributed Data (RDD) framework, which distributes the data frame into smaller chunks on different machines for faster computation. It also supports Map and Reduce for data processing. It supports SQL, data streaming, graph processing algorithms and machine learning algorithms. Though Spark can be accessed with Python, Java, and R, it has a strong support for Scala and is more stable at this point of time. It supports deep learning with sparkling water in H2O.

**Apache Hive:** is an open source platform that provides facilities for querying and managing large datasets residing in distributed storage (For example, HDFS). It is similar to SQL and it is called as HiveQL. It uses Map Reduce for processing the queries and also supports developers to plug in their custom mapper and reducer codes when HiveQL lacks in expressing the desired logic.

**Apache Pig:** is a platform that allows analysts to analyzing large data sets. It is a high-level programming language, called as Pig Latin for creating MapReduce programs that requires Hadoop for data storage. The Pig Latin code is extended with the help of User-Defined Functions that can be written in Java, Python and few other languages. It is amenable to substantial parallelization, which in turns enables them to handle very large data sets.

**Amazon Elastic Compute Cloud (EC2)**: is a web service that provides compute capacity over the cloud. It gives full control of the computing resources and allows developers to run their computation in the desired computing environment. It is one of the most successful cloud computing platform. It works on the principle of the pay-as-you-go model.

Few other frameworks that support big data are MongoDB, BlinkDB, Tachyon, Cassandra, CouchDB, Clojure, Tableau, Splunk and others.

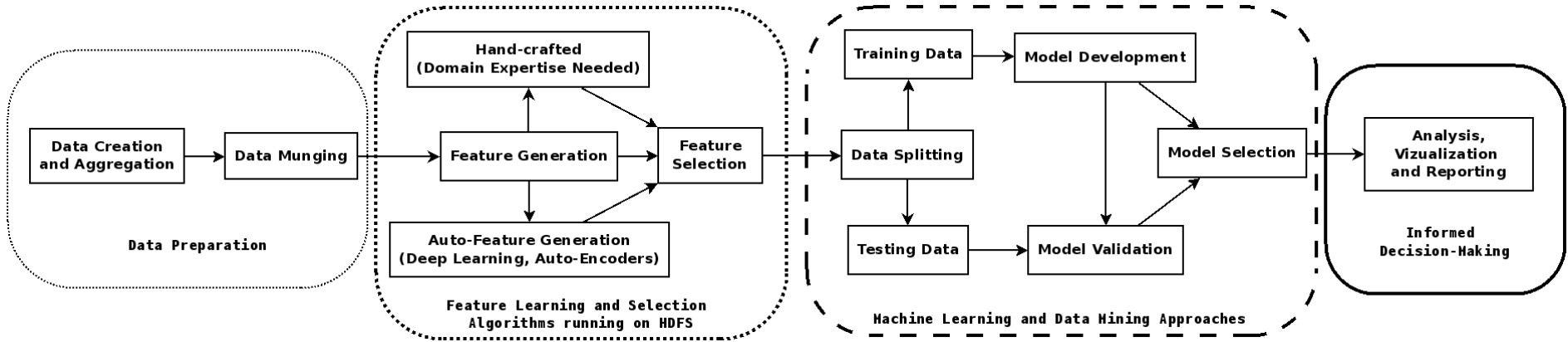

*Figure 4: Big Data Processes Illustration*

# APPLICATION OF BIG DATA ANALYTICS

The concept of big data analytics has left no sector untouched. Few sectors like Telecommunication, Retail and Finance have been early adopters of big data analytics, followed by other sectors (Villars et al., 2011). The application of big data analytics in various sectors is discussed as follows:

## Healthcare

Data analysts obtain and analyze information from multiple sources to gain insights. The multiple sources are electronic patient record; clinical decision support system including medical imaging, physician's written notes and prescription, pharmacy and laboratories; clinical data; and machine generated sensor data (Raghupathi and Raghupathi, 2014). The integration of clinical, public-health and behavioural data helps to develop a robust treatment system, which can reduce the cost and at the same time, improve the quality of treatment (Brown et al., 2011). Rizzoli Orthopedic Institute in Bologna, Italy analyzed the symptoms of individual patients to understand the clinical variations in a family. This helped to reduce the number of imaging and hospitalizations by 60% and 30%, respectively (Raghupathi and Raghupathi, 2014).

Obtaining information from external sources such as social media helps in early detection of epidemics and precautionary efforts. After the earthquake in Haiti in January 2010, analysis of tweets helped to track the spread of Cholera in the region (Raghupathi and Raghupathi, 2014). The data from the sensors are monitored and analyzed for adverse event prediction and safety monitoring (Mukherjee et al., 2012).

Artemis, a system developed by Blount et al. (2010), monitors and analyzes the physiological data from sensors in the intensive care units to detect the onset of medical complications, especially, in the case of neo-natal care. The real-time analysis of a huge number of claims requests can minimize fraud.

## Telecommunication

Low adoption of mobile services and churn management are few of the most common problems faced by the mobile service providers (MSPs). The cost of acquiring new customer is higher than retaining the existing ones. Customer experience is correlated with customer loyalty and revenue (Soares, 2012a,b). In order to improve the customer experience, MSPs analyze a number of factors such as demographic data (gender, age, marital status, and language preferences), customer preferences, household structure and usage details (CDR, internet usage, value-added services (VAS)) to model the customer preferences and offer a relevant personalized service to them. This is known as targeted marketing, which improves the adoption of mobile services, reduces churn, thus, increasing the revenue of MSPs. Ufone, a Pakistan-based MSP, reduced the churn rate by precisely marketing the customized offers to their customers (Utsler, 2013). The company analyzes the CDR data to identify the call patterns to offer different plans to customers. The services are marketed to the customers through a call or text message. Their responses are recorded for further analysis.

Telecom companies are working towards combating telecom frauds. Often, traditional fraud management systems are poor at detecting new types of fraud. Even they detect the occurrence of fraud lately, by then fraudsters would have changed their strategy. In order to overcome the limitations of traditional fraud management system, MSPs are analyzing real-time data to minimize the losses due to fraud. Mobileum Inc., one of the leading telecom analytics solution providers, is working towards providing a real-time fraud detection system using predictive analytics and machine learning (Ray, 2015).

Network Analytics is the next big thing in Telecom, where MSPs can monitor the network speed and manage the entire network. This helps to resolve the network problems within few minutes and helps to improve the quality of service and the customer experience. With the diffusion of Smartphones, based on analysis of real-time location and bevioural data, location-based services/context-based services can be offered to the customers when requested. This would increase the adoption of mobile services.

**Financial Firms**

Currently, capital firms are using advanced technology to store huge volumes of data. But increasing data sources like Internet and Social media require them to adopt big data storage systems. Capital markets are using big data in preparation for regulations like EMIR, Solvency II, Basel II etc, anti-money laundering, fraud mitigation, pre-trade decision-support analytics including sentiment analysis, predictive analytics and data tagging to identify trades (Verma and Mani, 2012). The timeliness of finding value plays an important role in both investment banking and capital markets, hence, there is a need for real-time processing of data.

**Retail**

Evolution of e-commerce, online purchasing, social-network conversations and recently location-specific smartphone interactions contribute to the volume and the quality of data for data-driven customization in retailing (Brown et al., 2011). Major retail stores might place CCTV not only to observe the instances of theft but also to track the flow of customers (Villars et al., 2011). It helps to observe the age group, gender and purchasing patterns of the customers during weekdays and weekends. Based on the purchasing patterns of the customers, retailers group their items using a well-known data mining technique called Market Basket Analysis (proposed by (Agrawal and Srikant, 1994)), so that a customer buying bread and milk might purchase jam as well. This helps to decide on the placement of objects and decide on the prices (Brown et al., 2011; Villars et al., 2011). Nowadays, e-commerce firms use market basket analysis and recommender systems to segment and target the customers. They collect the click stream data, observe behavior and recommend products in the real time.

Analytics help the retail companies to manage their inventory. For example, Stage stores, one of the brand names of Stage Stores Inc. which operates in more 40 American states, used to analytics to forecast the order for different sizes of garments for different geographical regions (Meek, 2015).

**Law Enforcement**

Law enforcement officials try to predict the next crime location using past data i.e., type of crime, place and time; social media data; drone and smartphone tracking. Researchers at Rutgers University developed an app called RTM Dx to prevent crime and is being used by police department at Illinois, Texas, Arizona, New Jersey, Missouri and Colorado. With the help the app, the police department could measure the spatial correlation between the location of crime and features of the environment (Mor, 2014).

A new technology called facial analytics that examines images of people without violating their privacy. Facial analytics is used to check child pornography. This saves the time of manual examination. Child pornography can be identified by integration of various technologies like Artemis and PhotoDNA by comparing files and image hashes with existing files to identify the subject as adult or child. It also identifies the cartoon based pornography (Ricanek and Boehnen, 2012).

## Marketing

Marketing analytics helps the organizations to evaluate their marketing performance, to analyze the consumer behavior and their purchasing patterns, to analyze the marketing trends which would aid in modifying the marketing strategies like the positioning of advertisements in a webpage, implementation of dynamic pricing and offering personalized products (Soares, 2012a).

## New Product Development

There is a huge risk associated with new product development. Enterprises can integrate both external sources, i.e., twitter and Facebook page and internal data sources, i.e., customer relationship management (CRM) systems to understand the customers' requirement for a new product, to gather ideas for new product and to understand the added feature included in a competitor's product. Proper analysis and planning during the development stage can minimize the risk associated with the product, increase the customer lifetime value and promote brand engagement (Anastasia, 2015). Ribbon UI in Microsoft 2007 was created by analyzing the customer data from previous releases of the product to identify the commonly used features and making intelligent decisions (Fisher et al., 2012).

## Banking

The investment worthiness of the customers can be analyzed using demographic details, behavioral data, and financial employment. The concept of cross-selling can be used here to target specific customer segments based on past buying behavior, demographic details, sentiment analysis along with CRM data (Forsyth, 2012; Coumaros et al., 2014).

## Energy and Utilities

Consumption of water, gas and electricity can be measured using smart meters at regular intervals of one hour. During this interval, a huge amount of data is generated and analyzed to change the patterns of power usage (Brown et al., 2011). The real-time analysis reveals energy consumption pattern, instances of electricity thefts and price fluctuations.

## Insurance

Personalized insurance plan is tailored for each customer using updated profiles of changes in wealth, customer risk, home asset value, and other data inputs (Brown et al., 2011). Recently, driving data of customers such as miles driven, routes driven, time of day, and braking abruptness are collected by the insurance companies by using sensors in their cars. Comparing individual driving pattern and driver risk with the statistical information available such as peak hours of drivers on the road develops a personalized insurance plan. This analysis of driver risk and policy gives a competitive advantage to the insurance companies (Soares, 2012a; Sun and Heller, 2012).

### Education

With the advent of computerized course modules, it is possible to assess the academic performance real time. This helps to monitor the performance of the students after each module and give immediate feedback on their learning pattern. It also helps the teachers to assess their teaching pedagogy and modify based on the students' performance and needs. Dropout patterns, students requiring special attention and students who can handle challenging assignments can be predicted (West, 2012). Beck and Mostow (2008) studied the student reading comprehension using intelligent tutor software and observed that reading mistakes reduced considerably when the students re-read an old story instead of a new story.

### Other sectors

With increasing analytics skills among the various organizations, the advantage of big data analytics can be realized in sectors like construction and material sciences (Brown et al., 2011).

## TECHNOLOGICAL GROWTH AND TECHNOLOGICAL LIMITATIONS

Advantages and applications of big data analytics are being realized in various sectors. The development of distributed file systems (eg., HDFS), Cloud computing (eg., Amazon EC2), in-memory cluster computing (eg., Spark), parallel computing (eg., Pig), emergence of NoSQL frameworks, advancement in machine learning algorithms (eg., Support Vector Machines, Deep Learning, Auto-Encoders, Random Forest) have brought big data processing a reality.

Despite the growth in these technologies and algorithms to handle big data, there are there are few limitations, which are discussed in this section.

1. **Scalability and Storage Issues:** The rate of increase in data is much faster than the existing processing systems. The storage systems are not capable enough to store these data (Chen et al., 2014; Li and Lu, 2014; Kaisler et al., 2013; Assunção et al., 2015). There is a need to develop a processing system that not only caters to today's needs but also future needs.
2. **Timeliness of Analysis:** The value of the data decreases over time. Most of the applications like fraud detection in telecom, insurance and banking, require real time or near real time analysis of the transactional data (Chen et al., 2014; Li and Lu, 2014).
3. **Representation of Heterogeneous Data:** Data obtained from various sources are heterogeneous in nature. Unstructured data like Images, videos and social media data cannot be stored and processed using traditional tools like SQL. Smartphones now record and share images, audios and videos at an incredibly increasing rate, forcing our brains to process more. However, the process for representing images, audios and videos lacks efficient storage and processing (Chen et al., 2014; Li and Lu, 2014; Cuzzocrea et al., 2011).
4. **Data Analytics System:** Traditional RDBMS are suitable only for structured data and they lack scalability and expandability. Though non-relational databases are used for processing unstructured data, but there exist problems with their performances. There is a need to design a system that combines the benefits of both relational and non-relational database systems to ensure flexibility (Chen et al., 2014; Li and Lu, 2014;).

5. **Lack of talent pool:** With the increase in amount of (structured and unstructured) data generated, there is a need for talent. The demand for people with good analytical skills in big data is increasing. Research says that by 2018, as many as 140,000 to 190,000 additional specialists in the area of big data may be required (Brown et al., 2011).

6. **Privacy and Security:** New devices and technologies like cloud computing provide a gateway to access and to store information for analysis. This integration of IT architectures will pose greater risks to data security and intellectual property. Access to personal information like buying preferences and call detail records will lead to increase in privacy concerns (Kaisler et al., 2013; Benjamins, 2014). Researchers have technical infrastructure to access the data from any data source including social networking sites, for future use whereas the users are unaware of the gains that can be generated from the information they posted (Boyd and Crawford, 2012). Big data researchers fail to understand the difference between privacy and convenience.

7. **Not always better data:** Social media mining has attracted researchers. Twitter has become a new popular source. Twitter users do not represent the global population. Big data researchers should understand the difference between big data and whole data. The tweets containing references to pornography and spam are eliminated resulting in the inaccuracy of the topical frequency. There is a redundancy in number of twitter users and twitter accounts one account accessed by multiple people and multiple accounts created by single user. There are active users and passive users who just sign in to listen. There are two types of accounts public and protected or private. Public tweets can be accessed through API. Few companies and start-ups have access to firehose (containing all public tweets posted and excludes private or protected tweets) while others have access to a gardenhose (roughly 10 percent public tweets) and a spritzer (roughly 1 percent of public tweets). Further, researchers have very limited access to firehose. Hence it does not provide the accuracy of the sample size of the dataset and interpretation obtained from analysis (Fisher et al., 2012; Boyd and Crawford, 2012, 2011).

8. **Out of Context:** Data reduction is one of common ways to fit into a mathematical model. Retaining context during data abstraction is critical. Data which are out of context lose meaning and value. There is an obsession for 'social graph' with the rise of social networking sites. Big data introduces two types of social networks: `articulated networks' and 'behavioural network'. Articulated networks are those resulting from specifying contacts through mediating technology. "Friends", "Acquaintances" in Facebook, "Follow" is twitter and "Best Friends", "Friends" and other circles in Google+ are examples of articulated network. Articulated networks are created to have separate group for friends, colleagues, friends of friends and filter the content that each group can view. Behavioural networks are obtained from social media interactions and communication patterns. But communication patterns necessarily need not reveal tie strength. Boss and a subordinate have large communication patterns compared to those with family members. Though the analysis of both articulated and behavioural networks reveals important insights but they are not equivalent to personal networks (Boyd and Crawford, 2012, 2011).

9. **Digital Divide:** Gaining access to big data is one of the most important limitations. Data companies and social media companies have access to large social data. Few companies decide who can access data and to what extent. Few sell the right to access for a high fees while others offer a portion of data sets to researchers. This results in "Digital Divide" in the realm of big data: Big Data rich and Big Data poor. There are three types of people and organization in this big data realm. First, people those who create data and this includes entire community who use web, mobile or other technologies. Second, people those who collect data and this group is small. Third, people who can analyze the data. This group is the smallest and the most privileged. Limited access to large social data explains the difficulty of conducting contemporary data-oriented social sciences and contemporary data-oriented humanities research (Boyd and Crawford, 2012, 2011; Manovich, 2011).
10. **Data errors:** With increase in the growth of information technology, huge amount of data is generated. With advent of cloud computing for storage and retrieval of data, there is a need to utilize the big data. Large datasets from internet sources are prone to errors and losses, hence unreliable. The source of the data should be understood to minimize the errors caused while using multiple datasets. The properties and limits of the dataset should be understood before analysis to avoid or explain the bias in the interpretation of data (Boyd and Crawford, 2011). For instance, consider the analysis of first-party and third-party pages in Social media, where the content in the first-party pages is verified whereas it is not the case in the latter (Kaisler et al., 2013).

## RESEARCH AREAS WITH TECHNOLOGICAL ENHANCEMENTS

Big data analytics is gaining so much attention these days but there are a number of research problems that still need to be addressed.

1) **Storage and Retrieval of Images, Audios and Videos:** Multidimensional data should be integrated with analytics over big data hence array-based in-memory representation models can be explored. Integration of multidimensional data models over big data requires the enhancement of query language HiveQL with multidimensional extensions (Cuzzocrea et al., 2011). With the proliferation of smart phones Images, Audios and Videos are being generated at an unremarkable pace. However, storage, retrieval and processing of these unstructured data require immense research in each dimension.

2) **Life-cycle of Data:** Most application scenarios require real-time performance of the big data analytics. There is a need to define the life cycle of the data, the value it can provide and the computing process to make the analytics process real time, thus, increasing the value of the analysis (Chen et al., 2014). Big data is always not always better, hence proper data filtering techniques can be developed to ensure correctness in the data (Boyd & Crawford, 2012). Another big issue is related to the availability of data that is complete and reliable. In most of the cases, data are sparse and do not show clear distribution, yielding misleading conclusions. A method to overcome these problems needs proper attention and sometimes handling of unbalanced data sets leads to biased conclusion (We et al., 2014).

3) **Big Data Computations:** Apart from current big data paradigms like Map-Reduce, other paradigms such as YarcData (Big Data Graph Analytics) and High-Performance Computing cluster (HPCC explores Hadoop alternatives), are being explored (Agneeswaran, 2012).

4) **Visualization of High-Dimensional Data:** Visualization aids in decision analysis at each and every step of the data analysis. Visualization issues are still part of data warehousing and OLAP research. There is a scope for visualization tools for high-dimensional data (Cuzzocrea et al., 2011; Chen et al., 2014; Li and Lu, 2014).

5) **Development of Algorithms for handling domain specific data:** Machine learning algorithms are developed to meet general requirements for processing data. However, it cannot replace the domain specific requirements and specific algorithms will be needed to gain insights from the desired discipline.

6) **Real time processing algorithms:** The pace at which data is being generated and the expectations from these algorithms may not be met, if the desired time delay is not met.

7) **Efficient storage devices:** The demand for storing digital information is increasing continuously. Purchasing and using available storage devices cannot meet this demand (Khan et al., 2014). Research towards developing efficient storage device that can replace the need for HDFS systems that is fault tolerant can improve the data processing activity and replace the need for software management layer.

8) **Social perspectives dimensions:** It is important to understand that any technology can yield faster results however, it is upto the decision makers to use it wisely. These results may have several social and cultural implications and sometimes leading to cynicism towards online platforms (Jothimani et al., 2015). There are few questions whether large-scale search data would help in creating better tools and services Or will it usher in privacy incursions and invasive marketing. Whether data analytics would help in understanding the online behavior, communities and political movements Or will it be used to track protesters and suppress freedom of speech (Boyd & Crawford, 2012).

## CONCLUSION

This chapter gave an overview of big data, processed involved in big data analytics and discussed various tools and techniques to process big data. We have also tried to compare different platforms for addressing big data storage, tools for handling big data, different libraries and packages have been highlighted. An overview of different languages used to handle big data has been covered. Different application domains where big data can play a significant role in improving the services have been discussed. Technological growth, limitations and direction for future research in improving big data have been highlighted.

## KEY DEFINITIONS

**Big Data**: refers to a huge amount of both structured and unstructured data that cannot be stored and analyzed using traditional database management techniques.

**Big Data Analytics**: is the process of obtaining, storing and analyzing high voluminous structured and unstructured data to obtain insights from the data.

**Cloud Computing**: is a process that provides infrastructure at a remote location using the Internet for storing and processing the data following a principle of model pay-as-you-go.

**Data Value Chain**: refers to the framework that deals with a set of activities to create value from the available data. It includes entire processes of data analysis including data generation, data collection, data transmission, data preprocessing, data storage and data analysis.

**Grid Computing**: is a distributed computing technology that allocates computing power that may require huge computational resources and is not available at the users' end. It works on a voluntary basis where users donate their memory and compute power of their computers so that other users can get benefitted.

**Hadoop**: is a Java-based framework that uses distributed computing environment to process a large amount of data.

**High Dimensional Data**: Data in which the number of variables is larger than the sample size.

**MapReduce**: is a software framework for processing a large dataset in a distributed manner in Hadoop clusters and uses Hadoop distributed file system.